# Designer polyradical nanographenes with strong spin entanglement and perturbation resilience *via* Clar's goblet extension

En Li[1,#], Manish Kumar[2, 3,#], Xinnan Peng[1,#], Tong Shen[1,#], Diego Soler-Polo[2], Yu Wang[1, 4], Yu Teng[1], Haoyu Zhang[1], Shaotang Song[1], Jishan Wu[1,*], Pavel Jelinek[2,*], Jiong Lu[1, 4]*

[1]Department of Chemistry, National University of Singapore, 3 Science Drive 3, Singapore 117543, Singapore

[2]Institute of Physics, Czech Academy of Sciences, Prague 16200, Czech Republic

[3]Department of Condensed Matter Physics, Faculty of Mathematics and Physics, Charles University, CZ12116 Prague 2, Czech Republic

[4]Institute for Functional Intelligent Materials, National University of Singapore, Singapore 117544, Singapore

[#]These authors contribute equally: En Li, Manish Kumar, Xinnan Peng, and Tong Shen

[*]E-mail: chmwuj@nus.edu.sg (J.W.); jelinekp@fzu.cz (P.J.); chmluj@nus.edu.sg (J.L.)

Abstract

**Polyradical nanographenes featuring strong spin entanglement and robust many-body spin states against external magnetic perturbations not only enable the exploration of correlated quantum magnetism at the molecular scale, but also constitute promising candidates for developing molecular qubits with chemical tunability and building scalable quantum networks. Here, we employed a predictive design strategy to achieve the on-surface synthesis of two homologues of Clar's goblet, $C_{62}H_{22}$ and $C_{76}H_{26}$, via lateral and vertical extensions of the parent structure, respectively. Vertical extension increases the number of topologically frustrated zero-energy modes, which scale linearly with the total number of benzene ring rows. In contrast, the lateral extension enhances electron-electron interactions, leading to the emergence of additional radical states beyond those predicted by the topological**

**zero-energy modes. Consequently, both structures exhibit correlated tetraradical character and a many-body singlet ground state as confirmed by multireference theoretical calculations. These magnetic states arise from unique magnetic origins and also display distinct resilience to external perturbations, which can be experimentally validated using nickelocene-functionalized scanning probe techniques. Our work presents a general strategy for rational design of highly entangled polyradical nanographenes with tunable spin numbers and resilience of their many-body spin states to perturbations, opening exciting possibilities for exploring novel correlated spin phases in molecular systems and advancing quantum information technologies.**

Open-shell graphene nanostructures have attracted tremendous interest as promising candidates for next-generation molecular spintronics and quantum computing[1, 2, 3]. Unlike the magnetism originating from localized *d*- or *f*- electrons in transition metals, organic radicals in carbon-based materials exhibit unique advantages, including negligible hyperfine interaction and spin−orbit coupling, which lead to long spin coherence times[4, 5]. In addition, the delocalized nature of π-electrons, combined with widely tunable spin configurations and magnetic exchange couplings[6, 7, 8, 9, 10, 11, 12, 13, 14, 15], makes these materials an ideal platform for realizing exotic quantum phases and developing molecular qubits for quantum technologies. Recent advances in on-surface synthesis and scanning probe microscopy have enabled the engineering of molecular nanomagnets with atomic precision and facilitated the exploration of various collective quantum states, including gapless excitations and the Haldane gap along with fractional edge states[16, 17, 18, 19, 20].

Among emerging carbon-based radicals, polyradical nanographenes with tunable multiple spins, strong spin entanglement, and resilience to external perturbation not only offer unique opportunities to explore exotic molecular quantum states[21], but also hold great promise as robust molecular qubits[22]. For instance, singlet-triplet (ST) transitions in correlated spin systems can be engineered to form decoherence-free subspaces, in which quantum states are insensitive to magnetic fluctuations and thus permit a long coherence time[23, 24]. In addition, the rational design of molecular building blocks enables the atomically precise assembly and integration of multiple π-spins into covalent sp² carbon networks[25, 26, 27, 28], which can host numerous electron-spin qubits that could potentially serve as a quantum simulator for correlated electron states. Although recent advances in on-surface synthesis have achieved the limiting case for such molecules[29, 30], there is a pressing need to establish a predictive strategy to guide the rational design of highly entangled polyradical nanographenes with tunable spin numbers and magnetic coupling strengths, especially in fully-fused open-shell nanographenes. Furthermore, the direct probing of their many-body spin states under local magnetic perturbations, crucial for the development of reliable spintronic devices and scalable molecular qubits, yet remains largely unexplored.

In this work, we employed a predictive design strategy to realize designer nanographenes with multiple correlated spins, tunable magnetic coupling strengths, and enhanced resilience to external perturbations. Through vertical or lateral extension of Clar's goblet, a prototypical topologically frustrated open-shell nanographene[6], we report the on-surface synthesis of two Clar's goblet homologues, $C_{62}H_{22}$ and $C_{76}H_{26}$, on Au(111). Scanning tunneling microscopy (STM) and non-contact atomic force microscopy (nc-AFM) were used to characterize the as-synthesized products at the single-chemical-bond level. Our inelastic electron tunneling spectroscopy (IETS) measurements, supported by theoretical calculations, reveal that both molecules exhibit a tetraradical character with strong multi-spin entanglement, yet arise from distinct magnetic origins associated with the lateral or vertical extensions. Furthermore, we employed a nickelocene-functionalized magnetic probe to assess the robustness of these designer polyradicals against magnetic perturbations, revealing different degrees of resilience.

**Design principles for polyradical nanographenes**
Clar's goblet features two unpaired electrons with a singlet ground state arising from topological frustration[6]. We explore designer polyradical nanographenes through the structural expansion of Clar's goblet. According to the hexagonal graphs theorem[31, 32], the number of zero-energy eigenstates in the tight-binding Hamiltonian for graphene nanoflakes is denoted by the term of nullity ($\eta$), which predicts the radical numbers in the absence of electron–electron (e-e) interactions. The nullity ($\eta$) is equal to the difference between the maximum numbers of non-adjacent vertices and edges. As illustrated in Figures 1a and S11, extending the structure of Clar's goblet vertically or laterally leads to distinct modulation of their radical numbers. A vertical extension leads to an increase in zero-energy modes, which scale linearly with (n−3), where n is defined as the total number of benzene ring rows. In contrast, lateral expansion consistently maintains invariant zero-energy modes, but increases the delocalization of molecular orbitals and reduces the highest occupied molecular orbital (HOMO)–lowest unoccupied molecular orbital (LUMO) gap due to a wider connection region. The resulting enhancement in e-e interaction may give rise to additional correlated radical states beyond those predicted by topological frustration. Thus, vertical and lateral extensions of Clar's goblet contribute to radical enhancement through distinct mechanisms. Understanding and controlling these structural extensions offers a tunable route for engineering the radical properties of π-conjugated nanostructures.

To demonstrate such effects of structural extension, we combined in-solution and on-surface synthesis methods to fabricate two Clar's goblet homologues (highlighted in Figure 1a) on Au(111) surfaces under ultra-high vacuum (UHV) conditions. Molecule **1** is designed by a lateral extension of Clar's goblet, yielding a structure with $\eta = 2$. In contrast, molecule **2** incorporates both lateral and vertical extensions, resulting in a larger $\eta = 4$. The calculated Hückel energy spectra (left panel of Figure 1b and right panel of Figure 1c) reveal two and four zero-energy modes in the laterally and vertically extended structures, respectively. Unlike vertical extension, lateral extension significantly enlarges the "neck" region connecting the top and bottom segments, which thus reduces the energy gap between $\varphi_1$ and $\varphi_2$ (Figure 1b) and enhances e-e interaction in molecule **1**. Therefore, lateral extension favors spin-symmetry breaking through strong e-e interaction, resulting in two more radicals in molecule **1**. As a result, both molecules can exhibit tetraradical character. However, these unpaired electrons originate from distinct mechanisms: In molecule **2**, they arise solely from topological frustration, while in molecule **1**, they result from both topological frustration and e-e interaction.

**On-surface synthesis of molecules 1 and 2**

To fabricate these two tetraradical molecules **1** and **2**, we designed the precursors **1'** and **2'**, where precursor **1'** features a [3]rhombene core with four 2,6-dimethylphenyl substituents attached at its zigzag edges and precursor **2'** consists of a fused triangulene dimer core and four 2,6-dimethylphenyl substituents positioned at its zigzag edges (Figure 2d). Both precursors were obtained through multi-step organic synthesis (see Supplementary section 1). The precursors were then separately deposited onto Au(111) under UHV conditions, where each precursor formed close-packed self-assembled islands (Figures S1a and S2a). Subsequent thermal annealing at 500 K gave rise to the target products **1** and **2** through cyclodehydrogenation and ring-closure reactions. Large-scale STM images of both products reveal fully or partially fused products, as well as the randomly connected oligomers (Figures S1b and S2c). High-resolution STM imaging of an individual molecule (Figures 2b and 2e) and nc-AFM imaging with a CO-functionalized tip[33, 34] confirmed the successful synthesis of both molecules **1** and **2** by resolving their overall uniform topography and internal bonding structures, as shown in Figure 2c and 2f, respectively. The relatively low yields of molecules **1** and **2** are likely due to their high chemical reactivity, leading to random intermolecular coupling (Figures S1b and S2c). In addition to the thermal-induced cyclodehydrogenation process, the synthesis of both molecules **1** and **2** can also be achieved through the atomic tip manipulation of partially fused molecules, following a protocol similar to that reported in our previous work[25].

**Electronic structures and spin excitation of molecules 1 and 2**

Differential conductance (d$I$/d$V$) spectroscopic measurements were performed to investigate the local electronic structures of both molecules **1** and **2**. The d$I$/d$V$ spectrum recorded at the corner of molecule **1** shows three prominent peaks at sample biases of −1.28 V, −0.26 V, and 0.94 V (Figure 3a). The spatial distributions of these electronic states are captured by the corresponding d$I$/d$V$ maps (Figure 3b), which show good agreement with theoretical d$I$/d$V$ images (Figure 3c) calculated using the multireference Dyson orbitals approach[35]. This method provides a rigorous framework for describing the electronic transitions involved in the scanning tunneling spectroscopy (STS) measurements, capturing many-body interactions and yielding spatially resolved positive/negative ionization resonance (PIR/NIR) peaks observed in the STS[36]. In addition, the d$I$/d$V$ spectrum taken at the corner of molecule **2** reveals several key features at sample biases of 1.0 V, 0.65 V, −0.2 V, and −0.6 V (Figure 3d). The spatial distributions of these states for molecule **2** are displayed in the corresponding d$I$/d$V$ maps (Figure 3e). These two molecules exhibit similar spatial distributions of electronic states: the d$I$/d$V$ maps acquired at −1.28 V for molecule **1** and at −0.6 V for molecule **2** reveal three bright lobes located at both top and bottom zigzag edges. In contrast, the maps at −0.26 V for molecule **1** and at −0.2 V for molecule **2** display a characteristic nodal pattern around four triangulene corners. The unoccupied states at 0.94 V for molecule **1** and at 1.0 V for molecule **2** exhibit considerably higher intensity in the central region, compared to those of occupied states. As shown in Figure 3f, the simulated constant-height d$I$/d$V$ maps derived from Dyson orbitals also show excellent agreement with the experimental data.

In addition to different ionic resonances, zoomed-in d$I$/d$V$ spectra in the vicinity of Fermi energy ($E_F$) reveal a symmetric step-like feature around $E_F$ for both molecules **1** and **2** (upper panels in Figures 4a and 4c), which can be attributed to spin-flip inelastic excitations from their ground state to an excited state. The excitation thresholds, extracted from the corresponding IETS spectra, appear at 30 meV for molecule **1** and 9 meV for molecule **2** (bottom panels in Figures 4a and 4c). To visualize the spatial distribution of these excitations, we recorded d$I$/d$V$ maps at −30 mV and −9 mV for molecules **1** and **2** (Figures 4b and 4d), respectively, both of which show excellent agreement with the calculated maps, as discussed in the following section.

**Theoretical investigations of molecules 1 and 2**

To gain deeper insights into the electronic and magnetic structure of these two molecules, we performed multireference calculations using the complete active space configuration interaction (CASCI) method. Both molecules **1** and **2** exhibit many-body wavefunctions consisting of multiple Slater determinants and adopt an open-shell singlet ground state, as indicated by the multireference ground state wavefunction coefficients shown in Figure

S14. Analysis of the occupation numbers of the natural orbitals constructed from the CASCI ground state wavefunction reveals that both molecules exhibit a tetraradical nature, with four unpaired electrons (Figures S5 and S6). The tetraradical character of molecule **2** arises from the presence of four topological zero-energy modes ($\eta = 4$), as previously discussed. In contrast, molecule **1**, despite having only two zero-energy modes ($\eta = 2$), also exhibits tetraradical character, as evidenced by its many-body natural orbitals (Figure S5). To elucidate the origin of this tetraradical behavior, we analyzed the one-electron tight-binding spectrum. As shown in Figure S11, the energy gap between the Hückel HOMO-1 and the LUMO+1 is smaller in molecule **1** (0.64 eV) than in molecule **2** (1.12 eV). The reduced HOMO-1 and LUMO+1 gap in molecule **1** is caused by the lateral expansion of Clar's goblet structure, which widens the central region connecting the two sets of sublattices. This widening leads to increased orbital delocalization of both the HOMO and LUMO in one-electron picture (Figure S12), effectively lowering the on-site Coulomb repulsion (U) for these orbitals. As a result of this smaller gap, e-e interactions become significant, promoting charge fluctuation from occupied to unoccupied frontier orbitals. This effect causes the emergence of two additional unpaired electrons beyond the two radicals associated with the zero modes ($\eta = 2$). This explains the tetraradical character observed in molecule **1**, which thus exhibits a total of four unpaired electrons. In contrast, the weaker e-e interactions in molecule **2** do not result in the formation of additional unpaired electrons.

The singlet-triplet excitation energy for molecule **1**, calculated using NEVPT2[37], is 44 meV, which closely matches the experimental value of 30 meV. For molecule **2**, the calculated energy difference between the first excited state and the ground state is 13 meV, also in good agreement with the experimental result of 9 meV. To simulate the d$I$/d$V$ maps corresponding to the IETS spin excitation maps, we computed the natural transition orbitals (NTOs)[38] for the spin-flip process between the ground state and the first excited state using the many-body CASCI wavefunctions. The simulated spin excitation maps for molecules **1** and **2**, obtained using NTOs, closely reproduce the experimental d$I$/d$V$ maps, as shown in Figures 4b and 4d, respectively. This strong correspondence confirms that the observed IETS features originate from spin-excitation processes.

The many-body CASCI calculations reveal that the open-shell singlet ground state of both molecules is characterized by a combination of ferromagnetic and antiferromagnetic exchange couplings between each pair of the four spins, resulting in a highly entangled multi-spin system. As illustrated in the conceptual spin correlation model in Figure 4f, spins localized on the same sublattices exhibit ferromagnetic coupling, while those localized on different sublattices exhibit antiferromagnetic coupling. The spin entanglement in both molecules is further supported by the calculated spin–spin correlation function $A_{ij} = \langle \hat{s}_i \hat{s}_j \rangle - \langle \hat{s}_i \rangle \langle \hat{s}_j \rangle$ between each pair of spins $i$ and $j$, as each spin is predominantly localized at one of the four corners of the molecule (Figures 4e and 4g).

**Magnetic resilience of molecules 1 and 2 under a nickelocene-functionalized tip**

To compare the magnetic resilience of the two highly entangled tetraradical molecules, we employed IETS measurements using a nickelocene ($NiCp_2$)-functionalized probe. The $NiCp_2$ molecule was chosen as an external magnetic perturbation, because the energy splitting between its ground state (S=1, $m_s$=0) and excited state (S=1, $m_s$=±1) is 4 meV[39,40], which closely matches the characteristic energy dissipation at liquid nitrogen temperature, $k_B T \ln(2) \approx 4.6\ meV$, where $k_B$ is Boltzmann's constant and T is the temperature[41]. After confirming the stability of the $NiCp_2$ tip (Figure S16), we recorded a set of IETS spectra above the molecules at different tip-sample distances, which alters the magnetic coupling strength between $NiCp_2$ and the probed molecule. As shown in Figures 5b and 5d, both molecules **1** and **2** display two sets of peak/dip features: The features at ±4 mV arise from the $NiCp_2$ excitation, while the features at ±34 mV for molecule **1** or ±14 mV for molecule **2** correspond to the coupling between each molecule and $NiCp_2$. As the tip approaches molecule **1**, the energetic positions of both IETS features remain nearly unchanged while their intensities increase. In sharp contrast, for molecule **2**, the peak and dip features at ±4 mV shift towards $E_F$, while the features at ±14 mV shift far away from $E_F$, accompanied by monotonic broadening, resembling the behavior observed in butterfly-shaped nanographene[29].

Although the two molecules exhibit the same spin-correlated ground state, they show different resilience to external magnetic perturbation by the NiCp$_2$ tip. To elucidate these differences, we employed a combination of the Heisenberg spin Hamiltonian and cotunneling theory[42, 43] to simulate the corresponding IETS spectra (see Supplementary section 7). As illustrated in the spin model in Figure 5a, decreasing the tip-sample distance increases the exchange interaction $J(z)$ between NiCp$_2$ and the molecule. In the weak coupling regime, the eigenvalues from the Heisenberg model predict three features: a peak at 4 mV (NiCp$_2$ excitation), a peak at 30 mV (molecule **1** excitation), and a third peak at 34 mV (simultaneous excitation of both NiCp$_2$ and molecule **1**), denoted as A, B, and C in Figure 5c and S15a, respectively. In the simulated IETS spectra with cotunneling theory, however, only the 4 mV and 30 mV peaks are observed. The 34 mV peak, associated with the joint excitation, is not visible due to selection rules imposed by cotunneling theory, which filters the observable transitions based on spin-selection rules and perturbative coupling to the electron reservoir (Figure 5c). As the NiCp$_2$ tip further approaches molecule **1**, both the molecular excitation (30 mV) and the joint excitation (34 mV) become increasingly visible as a result of the enhanced exchange coupling between the tip and the molecule. However, the excitations of NiCp$_2$ and molecule **1** remain well separated in energy. Due to this large energy gap of 30 meV, the spin states do not mix (see Figure S15c), with no renormalization of excitation energies. As a result, these states are resilient to perturbations as the exchange interaction increases.

The Heisenberg spectrum of molecule **2** reveals several eigenvalue transitions: 4 mV (NiCp$_2$ excitation), 9 mV (excitation of molecule **2**), 13 mV (joint excitation of NiCp$_2$ and molecule **2**), and 27 mV (second excitation of molecule **2** from singlet to quintet state), denoted as A', B', C', and D' in Figure 5d. Due to the same spin-selection rules from cotunneling theory, only the 4 mV and 9 mV transitions appear in the IETS signal. The joint excitation at 13 mV and the higher-energy transitions at 27 mV are suppressed and not observed in the weak coupling regime. As the tip approaches molecule **2**, the exchange interaction strengthens, leading to mixing between the singlet and triplet states of molecule **2** with the NiCp$_2$ excitation (Figure S15d). This interaction results in a shift of the 4 mV peak toward zero bias, while the 9 mV peak shifts to larger bias, eventually merging with the joint excitation feature of NiCp$_2$ and molecule **2** (Figure 5e). The absence of a signal at 27 mV, despite its presence in the Heisenberg model, is clearly explained by the spin-selection rules of IETS, which forbid this transition from ground state to second excited state. Consequently, no corresponding peak is observed at larger bias.

Thus, the substantial differences in excitation energies between the two molecules result in their distinct responses to external magnetic perturbations from nickelocene, as illustrated in Figure 1c. For molecule **1**, which has a large excitation energy of 30 meV, the magnetic excitation of NiCp$_2$ remains decoupled from the molecular excitation, indicating negligible interaction between the two systems. In contrast, for molecule **2**, the energy levels of the molecule and nickelocene are sufficiently close (9 meV vs 4 meV) to enable state mixing, resulting in hybridized excitations. The distinct resilience of the two molecules enables them to function as spin sensors or robust quantum bits.

In summary, we have demonstrated a comprehensive framework for engineering highly entangled polyradical nanographenes with tetraradical character through topological expansion of Clar's goblet. By applying both lateral and vertical structural modifications, we reveal distinct mechanisms of radical generation: lateral extension enhances electron correlation and orbital delocalization, while vertical extension increases the number of topological zero-energy modes. Our findings provide a powerful and predictive guideline for the rational design of polyradical nanographenes with tunable spin numbers. Moreover, the two molecules display distinct responses to a magnetic NiCp$_2$ probe due to their different excitation energies, opening new opportunities for the development of perturbation-resilient qubits and molecule-based spin sensors. The realization of highly entangled polyradical nanographenes with strong magnetic robustness paves the way for advances in quantum information technologies.

During the preparation of this manuscript, we have become aware of a related work[44] on synthesis and characterization of π-extended Clar's goblet ($C_{76}H_{26}$).

## Methods

### On-surface synthesis

All experiments were carried out in low-temperature scanning tunneling microscopy/atomic force microscopy (LT–STM/AFM) systems operated under ultrahigh vacuum (base pressure <2.0 × 10$^{-10}$ mbar) at temperatures of 4.5 K (Scienta Omicron) and 5.0 K (CreaTec Fischer), respectively. The Au(111) single-crystal surface was prepared through cycles of Ar+ sputtering (0.8 keV) and subsequent annealing at 770 K. Powder samples of precursor **1'** were sublimed onto Au(111) from a Knudsen cell at 560 K, while precursor **2'** was sublimed at 620 K. After further annealing, the sample was transferred to a cryogenic scanner for characterization.

### STM/STS and nc-AFM measurements

Two qPlus sensors were used for nc-AFM imaging. One is sensor S0.8[45] with a resonance frequency of $f_0$ = 39.646 kHz, a stiffness of 3600 N/m, and a quality factor of 119 K, while the other one is a commercial sensor with a resonance frequency of $f_0$ = 28 kHz, a stiffness of 1800 N/m, and a quality factor of 16 K. nc-AFM images were collected in constant-height frequency modulation mode using an oscillation amplitude of A = 40-100 pm. Differential conductance (d$I$/d$V$) spectra were collected using the standard lock-in technique with a frequency of 737.3 Hz. The amplitude of modulation was 20mV for molecular orbital measurements and 1 mV for spin excitation measurements, respectively. STM images and d$I$/d$V$ maps were recorded in constant-current mode, while d$I$/d$V$ spectra were recorded in constant-height mode.

The measurements with a nickelocene-functionalized tip were conducted at 5.0 K. The nickelocene molecules were dosed onto the Au(111) surface below 10 K. Before conducting spectroscopy measurements on the molecules, we performed height-dependent measurements on the bare Au(111) surface, to confirm the stability of NiCp$_2$ tip.

**Theoretical calculations**

The molecular geometries were optimized using density functional theory (DFT), as implemented in the FHI-AIMS software package[46] employing the PBE0 hybrid functional[47]. In these calculations, the Tkatchenko-Scheffler method was used to account for van der Waals interactions[48]. Because of the multi-radical nature of the molecules, the complete active space configuration interaction (CASCI) method was employed to obtain an accurate description of the wave function and electronic energies using home-built code. One- and two-electron integrals were constructed in the basis of molecular orbitals in an active space of 12 electrons in 12 orbitals [CAS(12,12)] around the Fermi energy, and were calculated using the quantum chemistry software ORCA[49] using the orbitals obtained from the DFT-PBE[50]. To determine the number of unpaired electrons in the molecule, we calculated the occupation of natural orbitals by diagonalization of the one-particle density matrix constructed from the ground state many-body CASCI wavefunction.

For further correct for out-of-CAS dynamical electron correlation, we employed the Quasidegenerate Second-Order N-Electron Valence State Perturbation Theory (QD-NEVPT2)[37] to calculate the excitation energy of the molecule.

To interpret the d$I$/d$V$ maps in STM for electron addition and removal processes, the many-body Dyson orbitals[36] were constructed from the CASCI wavefunction. To simulate the dI/dV maps corresponding to the IETS spin excitation maps, we calculated the natural transition orbitals (NTOs)[38] for the spin-flip process between the ground state and the first excited state using the many-body CASCI wavefunctions. Theoretical dI/dV maps were calculated using the Probe Particle Scanning Probe Microscopy (PP-SPM) code[51, 52] for a metal-like tip without tip relaxation.

## Acknowledgements

J.L. acknowledges support from the NRF, Prime Minister's Office, Singapore, under the Competitive Research Program Award (NRF-CRP29-2022-0004) and MOE grants (MOE-T2EP10221-0005, MOE-T2EP10123-0004, MOE-T2EP10223-0004 and MOE-T2EP10124-0004). S.S. acknowledges the support from A*STAR under its AME YIRG Grant (M22K3c0094). P.J. appreciate financial support from the CzechNanoLab Research Infrastructure supported by MEYS CR (LM2018110) and GACR 23-05486S. J.W. acknowledges financial support from the Singapore Ministry of Education under a Tier 3 program (MOE-000755-00).

## Competing interests

The authors declare no competing interests.

## Author contributions

J.L. supervised the project and coodinated the collaborations. E.L., J.W. and J.L. conceived and designed the experiments. P.J. conceived the theoretical studies. E.L. and X.P. performed the on-surface synthesis and STM/STS and nc-AFM measurements. T.S. and J.W. synthesized the organic precursors. M.K. and D.S.-P. carried out theoretical calculations. S.S., Y.W., and H.Z. assisted with the data analysis and manuscript preparation. Y.T. helped with the data presentation. E.L., M.K., and J.L. wrote the manuscript with contributions from the other authors.

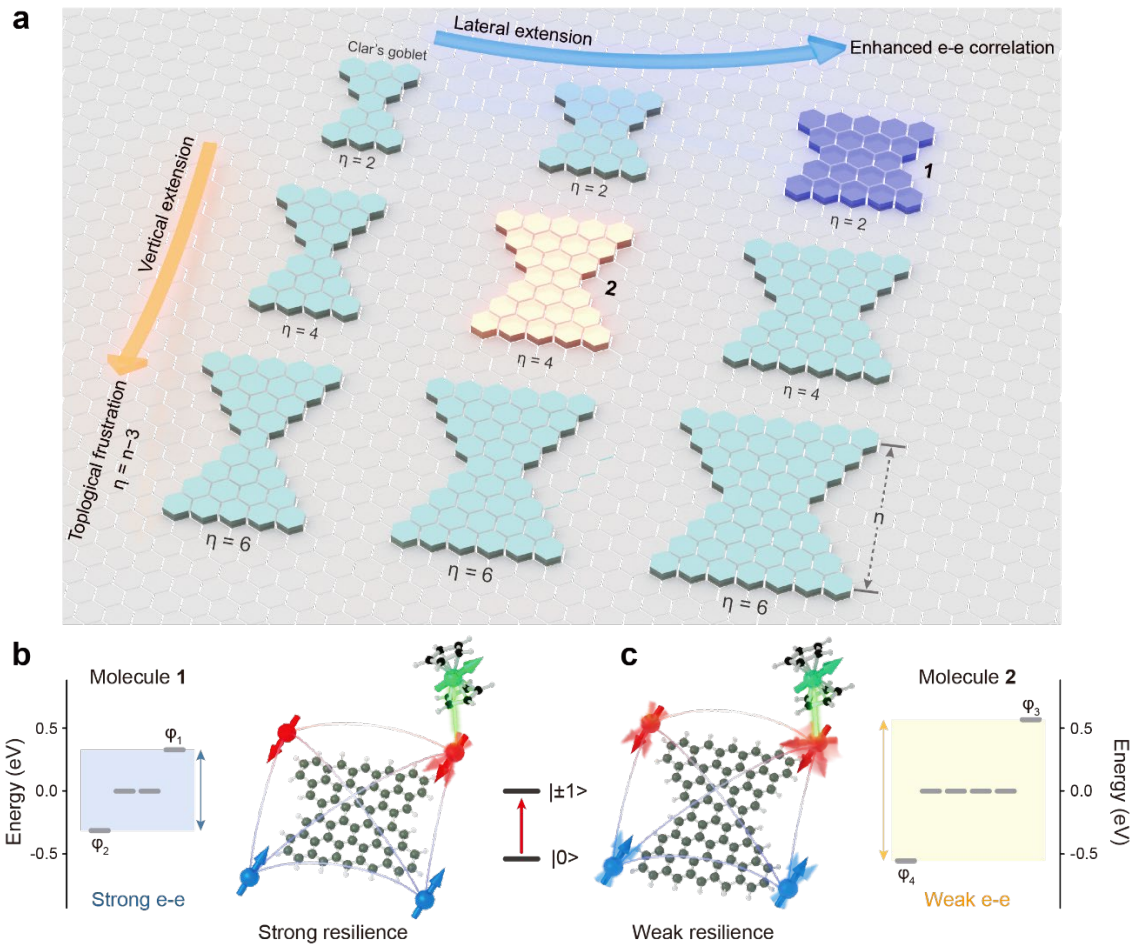

**Figure 1. Designer polyradical nanographenes through Clar's goblet extension.** (a) Illustration of open-shell nanographenes with different sizes by lateral and vertical extensions of Clar's goblet. The nullity (η) accounts for the number of zero-energy eigenstates in the tight-binding Hamiltonian for a nanographene. n denotes the total number of benzene ring rows. (b) Left: The Hückel energy spectrum of molecule **1**. A double-headed arrow indicates the energy gap between $\varphi_1$ and $\varphi_2$. Right: Illustration of spin correlation models for molecule **1** and its resilience to external magnetic perturbations from a nickelocene magnetic probe. (c) Left: Illustration of spin correlation models for molecule **2** and its weak resilience to nickelocene perturbations. Right: The Hückel energy spectrum of molecule **2**. The double-headed arrow indicates the energy gap between $\varphi_3$ and $\varphi_4$.

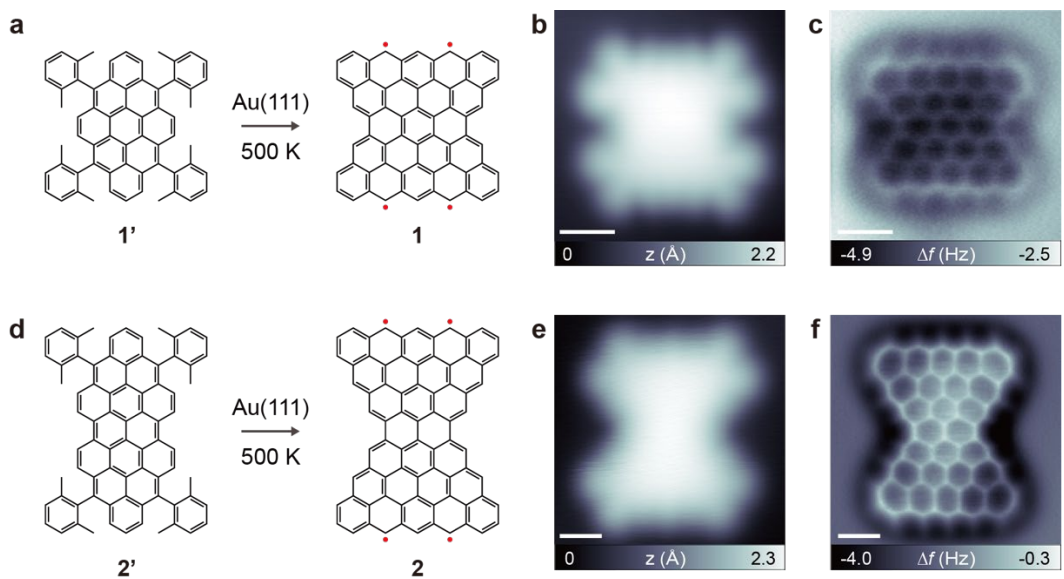

**Figure 2. On-surface synthesis of molecules 1 and 2.** (a) Illustration of the surface-assisted transformation of precursor **1'** to molecule **1**; (b) STM image (−0.3 V, 300 pA) of product **1**. (c) nc-AFM frequency shift image of molecule **1** captured using a CO-functionalized tip (resonant frequency: 28 kHz, oscillation amplitude: 100 pm). (d) Illustration of the on-surface synthetic strategy for molecule **2**. (e) STM image (−0.62 V, 400 pA) of product **2**. (f) nc-AFM frequency shift image of molecule **2** captured using a CO-functionalized tip (resonant frequency: 39.6 kHz, oscillation amplitude: 50 pm). STM images were taken using a metallic tip. Scale bar in (b), (c), (e), and (f): 0.5 nm.

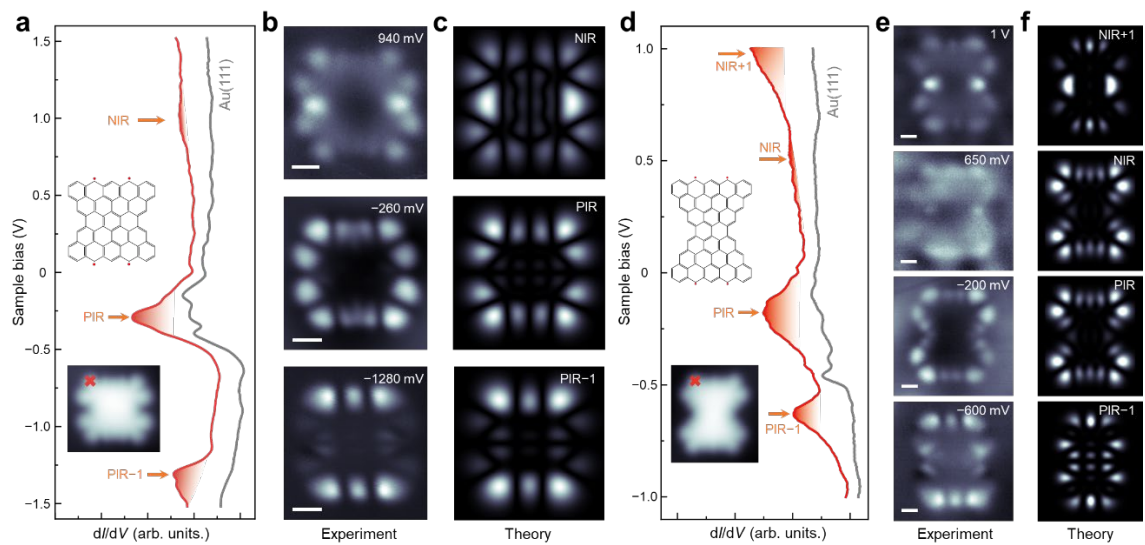

**Figure 3. Electronic structures of molecules 1 and 2.** (a) d$I$/d$V$ spectra acquired at molecule **1** and the Au(111) substrate (setpoint: −0.3 V, 500 pA). (b) Constant-current dI/dV maps taken at 0.94 V, −0.26 V, and −1.28 V. (c) Simulated dI/dV maps of Dyson orbitals for molecule **1**. (d) d$I$/d$V$ spectra obtained at molecule **2** and the Au(111) substrate (setpoint: −1.0 V, 800 pA). (e) dI/dV maps acquired at 1.0 V, 0.65 V, −0.2 V, and −0.6 V. (f) Simulated d$I$/d$V$ maps of Dyson orbitals for molecule **2**. The spectra in (a) and (d) have been shifted vertically for clarity. All spectra and STS maps were taken using a metallic tip. Scale bar in (b) and (e): 0.5 nm.

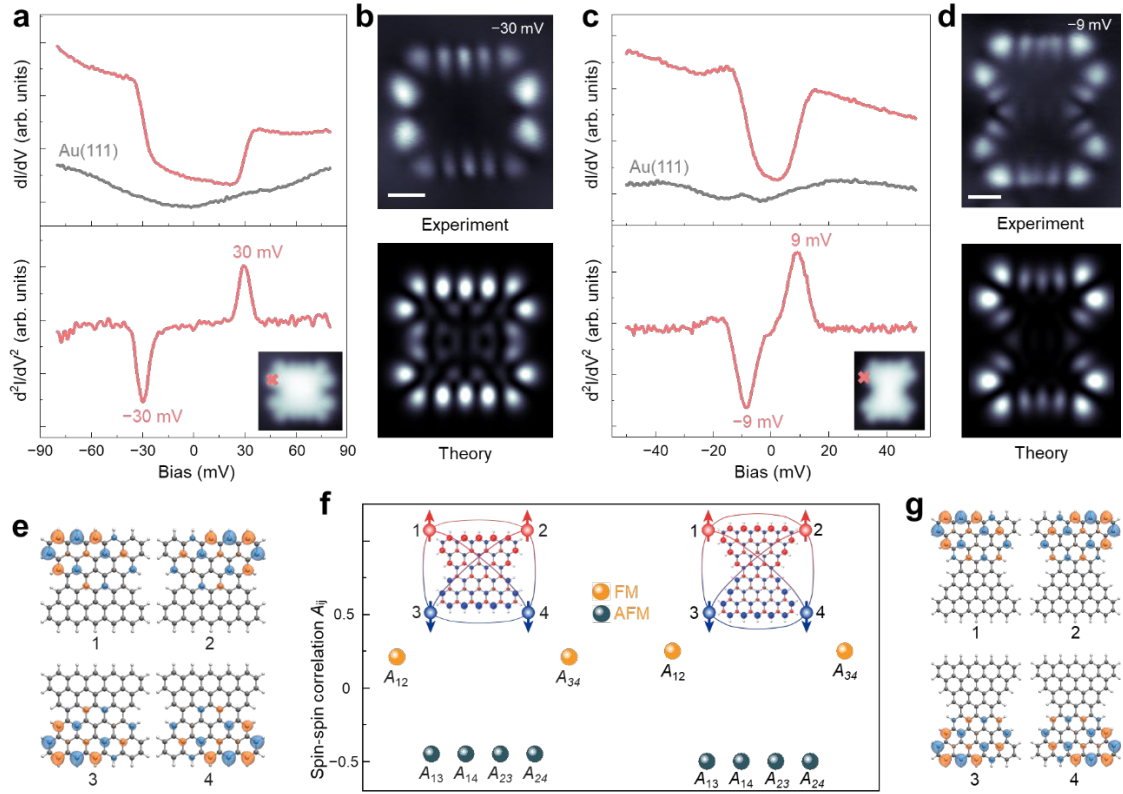

**Figure 4. Spin excitation and correlation in molecules 1 and 2. (**a) dI/dV spectra and IETS for molecule **1**, measured at the position marked by the red cross (inset). (b) Top: Experimental spin-excitation map of molecule **1** at –30 mV. Scale bar: 0.5 nm. Bottom: Simulated map of molecule **1** based on the sum of singlet–triplet CAS NTOs. (c) dI/dV spectra and IETS for molecule **2** at the position marked by the red cross (inset). (d) Top: Experimental spin-excitation map of molecule **2** at –9 mV. Bottom: Simulated map of molecule **2** based on the sum of singlet–triplet CAS NTOs. (e, g) Molecular orbitals showing spin localization at each corner for molecules **1** and **2**, respectively. (f) Spin–spin correlation for molecules **1** and **2**, calculated as $A_{ij} = \langle \hat{s}_i \hat{s}_j \rangle - \langle \hat{s}_i \rangle \langle \hat{s}_j \rangle$. Insets: Calculated spin density maps with conceptual spin interaction illustrations (blue/red: spin-up/spin-down). The spectra in (a) and (c) have been vertically offset for clarity. The spectra and STS maps were acquired using a metallic tip.

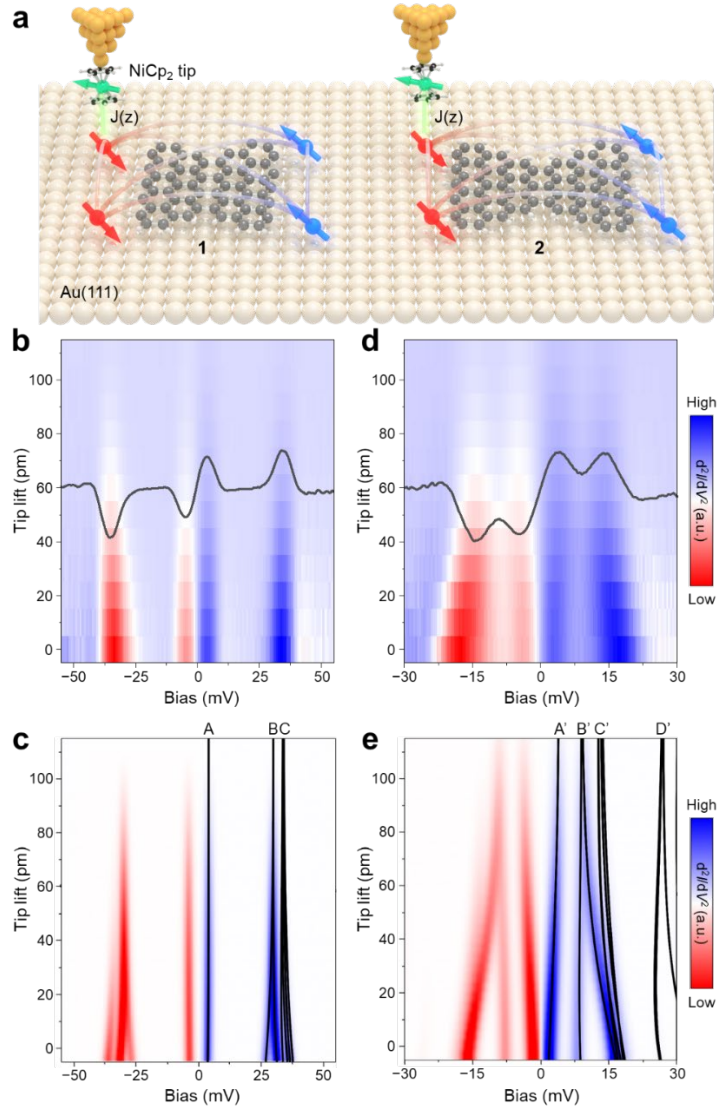

**Figure 5. Magnetic characterization of molecules 1 and 2 with a nickelocene-functionalized tip.** (a) Illustration of the measurement setup of molecules **1** and **2** with a NiCp$_2$ tip. (b) Height-dependent map composed of a series d$^2$I/dV$^2$ spectra acquired with a NiCp2 tip on molecule **1**. The overlaid black curve shows one of the IETS spectra acquired with the NiCp$_2$ tip. (c) Simulated IETS spectra for molecule **1** as a function of the coupling strength $J(z)$. A, B, and C originate from the coupling of $S_{z,Ni} = \pm 1$ with $S_{M1} = 0$, $S_{z,Ni} = 0$ with $S_{M1} = 1$, and $S_{z,Ni} = \pm 1$ with $S_{M1} = 1$, respectively. Similarly, (d) and (e) for the molecule **2**. A', B', C', and D' originate from the coupling of $S_{z,Ni} = \pm 1$ with $S_{M2} = 0$, $S_{z,Ni} = 0$ with $S_{M2} = 1$, $S_{z,Ni} = \pm 1$ with $S_{M2} = 1$, and $S_{z,Ni} = 0$ with $S_{M2} = 2$, respectively.